\documentclass[aps,pre,twocolumn,
floats,amsmath,amssymb,showpacs,showkeys,preprintnumbers,amsmath,amssymb,floatfix]%
{revtex4}

\usepackage{graphicx}
\usepackage{dcolumn}
\usepackage{color}
\usepackage{subfigure}

\begin{document}

\title{Convergent Chaos}

\author{Marc Pradas$^{2,1}$, Alain Pumir$^{3,1}$, Greg Huber$^1$ and Michael Wilkinson$^{2,1}$}

\affiliation{\\ 
$^1$ Kavli Institute for Theoretical Physics, University of California Santa Barbara, CA 93106, USA\\
$^2$ Department of Mathematics and Statistics,
The Open University, Walton Hall, Milton Keynes, MK7 6AA, England,\\
$^3$ Laboratoire de Physique,
 Ecole Normale Sup\'erieure de Lyon, CNRS, Universit\'e de Lyon,
 F-69007, Lyon, France,
}

\begin{abstract}
Chaos is widely
understood as being a consequence of sensitive dependence 
upon initial conditions. This is the result of an instability
in phase space, which separates trajectories exponentially.
Here, we demonstrate that this criterion should be refined.
Despite their overall intrinsic instability, trajectories may be very strongly
convergent in phase space over extremely long periods, as revealed by
our investigation of a simple chaotic system (a realistic model for
small bodies in a turbulent flow).
We establish that this strong convergence is a
multi-facetted phenomenon, in which the clustering is intense, widespread and
balanced by lacunarity of other regions.
Power laws, indicative of scale-free features, characterize
the distribution of particles in the system. 
We use large-deviation and extreme-value statistics to explain the effect. 
Our results show that the interpretation of the \lq butterfly 
effect' needs to be carefully qualified. We argue that the 
combination of mixing and clustering processes makes our specific model 
relevant to understanding the evolution of simple organisms. Lastly, this 
notion of \emph{convergent chaos}, which implies the existence of conditions 
for which uncertainties are unexpectedly small, may also be relevant
to the valuation of insurance and futures contracts.  
\end{abstract}

\pacs{05.40.-a,05.10.Gg,05.40.-a}
\keywords{chaos, Lyapunov exponent, futures contracts}

\maketitle

\section{Introduction}
\label{sec: 1}

The concept of \lq chaos' is one of the most salient paradigms of modern
science~\cite{Ott02}. The significance of the central notion
of exponential sensitivity to initial conditions
is emblematically illustrated by the \lq butterfly effect'.
The question ``Does the flap of a butterfly's wings in Brazil set off a tornado in Texas?"
was famously posed by E. N. Lorenz in a conference talk in 1972
\cite{Lor72,Lor95,Pal+14}.  In the roughly half century since Lorenz's work,
his question has invariably been conflated with: ``Can the flap of a butterfly's wings ... ?"
The affirmative answer to {\it that} question has cemented sensitive dependence
on initial conditions as a hallmark of chaotic systems, the weather system included.
But a deep, outstanding question behind the butterfly effect lies in Lorenz's
original formulation: Are perturbations destined to alter the course of large-scale
patterns in turbulent systems?  Or could regions of the phase space of a chaotic
dynamical system be screened off from small perturbations?  This is the real
import of Lorenz's Brazilian butterfly, and we note that Lorenz never definitively
answered his original question.

\begin{figure}
\centering
\includegraphics[width=0.49\textwidth]{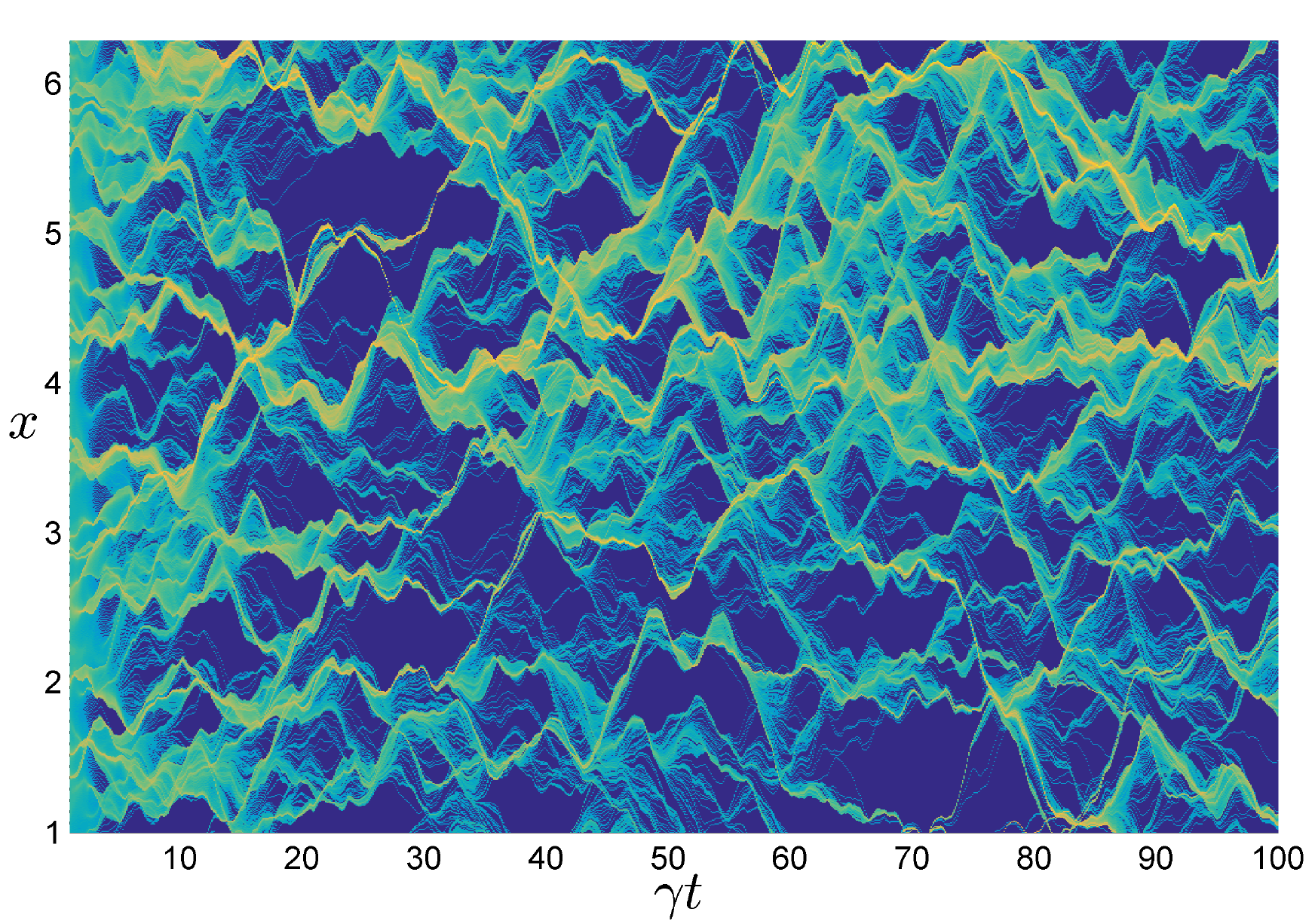}
\caption{
Trajectories, $x(t)$, for the dynamical system described by 
\eqref{eq: 1}: many trajectories show strong and long lasting convergence, despite the fact that
they must ultimately diverge. The colormap is chosen so that blue and yellow correspond to sparse and highly dense regions, respectively. Parameter values are quoted in the text. 
Clusters of trajectories can persist over durations long compared to the
expected separation time, which, with the model used here, is 
$\approx 15/\gamma$.
}
\label{fig: 1}
\end{figure}

It is the purpose of this paper to suggest an important and widely applicable
refinement of the concept of chaos, based upon results illustrated by figure \ref{fig: 1}.
This shows trajectories of particles in a model for the motion of particles in a turbulent
fluid flow. In order to simplify the discussion we consider a one-dimensional model, where
the position of a particle is $x(t)$ at time $t$.
In this model, it has been proven that trajectories separate
exponentially. In technical terms, the rate of separation of
trajectories (the Lyapunov exponent\cite{Ott02}) is positive. However, the trajectories illustrated in
figure \ref{fig: 1} show a strong tendency to cluster together, despite the fact that they
must eventually diverge.

In several one-dimensional chaotic systems, it has been observed that 
trajectories may show a temporary convergence preceding their eventual separation  
(see, for example \cite{Fuj83}, \cite{Aur+96}). 
Figure \ref{fig: 1} reveals that the convergence can
lead to clusters of trajectories, over times which are much longer than
the expected divergence time. 
Additionally, figure \ref{fig: 1} reveals that the simulated trajectories 
tend to form surprisingly dense clusters.
Quantitatively, for over $50\%$ of the time, $10\%$ of the
$1.5 \times 10^4$ trajectories used in figure \ref{fig: 1} are
clustered into a region of width $\Delta x = L/4000\approx 10^{-3}$, 
where $L=2\pi$ is the domain size. At some instants, up to $70\%$ 
of the total number of trajectories can accumulate in a region of size 
$L/4000$. 

Thus, the phenomenon illustrated in Figure \ref{fig: 1} indicates that,
despite the intrinsic unpredictability of the system on very long time scales,
there may be basins in the space of initial conditions which attract a significant
fraction of the phase space over a finite time, giving a final position
which is highly insensitive to the initial conditions. If the initial conditions which
are of physical interest lie within one of these basins, the behaviour of the system can be computed
accurately for a time which is many multiples of the inverse of the Lyapunov
coefficient. The possibility of the butterfly effect is contained in the
definition of chaos. Our results, however, indicate that the standard
definition of chaos, dependent upon a positive Lyapunov exponent, does not \emph{necessarily} imply a sensitive dependence upon initial conditions in practical applications, where we are only
concerned with finite times.

In this paper we discuss various quantitative aspects of the clustering effect
shown in figure \ref{fig: 1}. After introducing a canonical model in section \ref{sec: 2}, we give
a summary of our results on the strength of the effect (section \ref{sec: 3}). 
When we examine the structure of the patterns in figure \ref{fig: 1} statistically,
we find (section \ref{sec: 4}) that power-law relations are ubiquitous, indicating 
scale-free behaviour with universal characteristics~\cite{Strogatz:2001}. In section 
\ref{sec: 5} we explain strong convergence effect quantitatively by considering the 
finite-time Lyapunov exponent. Using a combination of large-deviation and extreme-value statistics approaches, we have been able to show that the minimum value of the finite-time Lyapunov exponent can remain negative for a very long time. 
In section \ref{sec: 6} we argue that some trajectories may show perpetually convergent 
behaviour. The phenomena described in our studies are expected to be realised in a wide
range of physically relevant models, and section \ref{sec: 7} discusses possible areas of 
application. 

\section{A simple chaotic system}
\label{sec: 2}

To stress the notion of intrinsic stochasticity in dynamical systems,
the most intensively studied models for chaos are purely deterministic.  
For the purpose of understanding generic physical 
processes, however, these models may lead to the physically artificial
situation where large regions of phase space are inaccessible at long time. 
In many extended physical systems, some degrees of freedom play a minor role,
and can be modelled stochastically. 
In addition, dynamical models that contain random elements are less prone
to lead to empty regions of phase space. These considerations 
provides a strong physical motivation to consider 
a dynamical model with random elements. 
In such a model, the emergence of sparse regions in
phase space, as found e.g. in the case of inertial particles in 
turbulent flows~\cite{Eaton+91},
necessarily results from a nontrivial dynamical property of the system. 

We therefore propose to consider a model in which the trajectories have a 
continuous dependence upon the phase point, but where the dynamics 
contains 
random elements. In order to eliminate irrelevant details, it is also 
desirable to have 
a model for which statistics of the phase-space velocity are invariant under translations
in time and space.

Among many possible abstract dynamical systems containing stochastic 
processes which satisfy these criteria, we have chosen a model which 
has a very direct physical interpretation, and which has already 
been extensively studied \cite{Fal+00}.
The model that we consider is a realistic description of a ubiquitous physical
phenomenon, namely the motion of small particles in a turbulent fluid.
The equations of motion are \cite{Max+83,Gat83}
\begin{eqnarray}
\label{eq: 1}
\dot x&=&v,
\nonumber \\
\dot v&=&\gamma[u(x,t)-v].
\end{eqnarray}
Here $\gamma$ is a constant describing the rate of damping
of motion of a small particle relative to the fluid and $u(x,t)$
is a randomly fluctuating velocity field of the fluid in which the
particles are suspended. In figure \ref{fig: 1} we solved  \eqref{eq: 1} 
on the interval $[0,L]$ with periodic boundary conditions, and a velocity
field where the correlation function is white noise in time, satisfying
$\langle u(x,t)\rangle=0$ and $\langle u(x,t)u(x',t')\rangle=\delta(t-t')C(x-x')$,
where angular brackets denote averages throughout. The correlation function
is $C(\Delta x)=\epsilon^2\xi^2\gamma\,\exp\left(-\Delta x^2/2\xi^2\right)$,
where $\xi$ is the correlation length and $\epsilon$ is a coupling constant.
The numerical parameters were $L=2 \pi$,
$\xi=0.08$, $\gamma=0.0112$ and $\epsilon=1.25\,\epsilon_c$,
where $\epsilon_c\approx 1.331$ is the value above which the
Lyapunov exponent becomes positive\cite{Wil+03}. 

The results obtained with model \eqref{eq: 1} for a simple $1$-dimensional 
system will be corroborated qualitatively by the results of a model
of a compressible $2$-dimensional flow, presented in Section \ref{sec: 7}. 

\begin{figure}[h t b]
\includegraphics[width=0.48\textwidth]{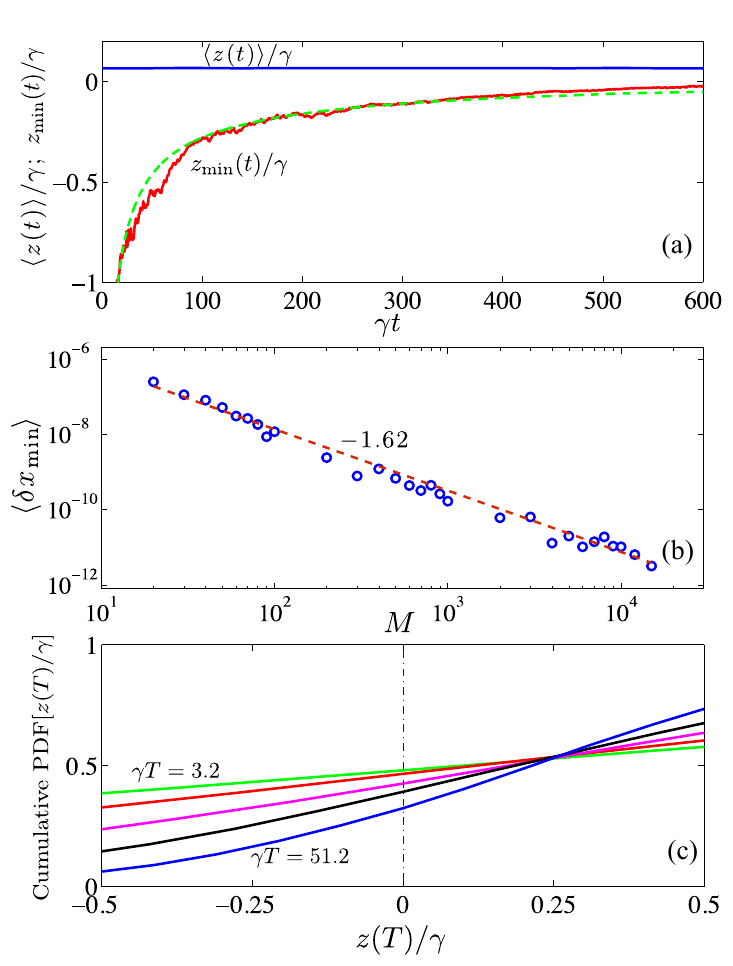}
\caption{(a) The minimum of the finite time Lyapunov exponent over $M$ trajectories
can remain negative, indicating converging trajectories, for very long times. It does
converge to the mean Lyapunov exponent $\lambda = \langle z(t)\rangle$ (positive for a chaotic system), but the
convergence described by \eqref{eq: 3} is very slow. The characteristic time
of trajectory separation is  $\gamma/\lambda \approx 15$. The exponent of the power
law fit (dashed line) is $\alpha \approx 0.6$. (b) The smallest separation between $M$ initially uniformly distributed  trajectories satisfied $\delta x_{\rm min}\sim M^{-\Gamma}$ with $\Gamma\approx 1.6$. (c)  Cumulative probability for the value of the finite-time
Lyapunov exponent, $z(t)$, at different values of the time (in dimensionless units).
The distribution of $z(t)$ is very broad, even for large values  of $t$. In all panels the
parameters are the same as for Figure \ref{fig: 1}.
}
\label{fig: 2}
\end{figure}

\section{Characterising the strong convergence of trajectories.}
\label{sec: 3}

The finite-time Lyapunov exponent (FTLE) at time $t$ for a trajectory
starting at $x_0$ is defined by \cite{Ott02}
\begin{equation}
\label{eq: 2}
z(t)=\frac{1}{t}\ln\,\left\vert\frac{\partial x_t}{\partial x_0}\right\vert_{x(0)=x_0}
\ ,
\end{equation}
where $x_t$ denotes position at time $t$.
\begin{figure*}[h t b]
\includegraphics[width=0.47\textwidth]{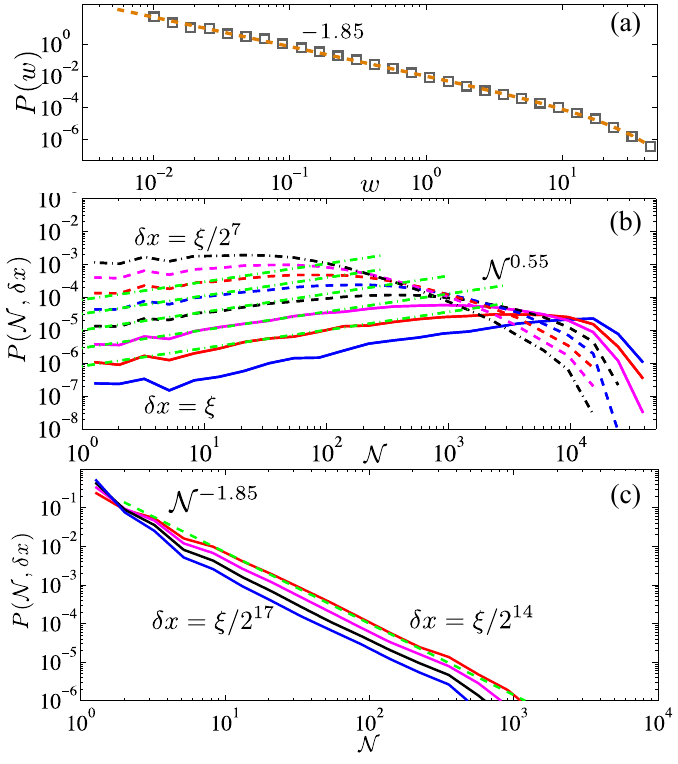}\
\includegraphics[width=0.49\textwidth]{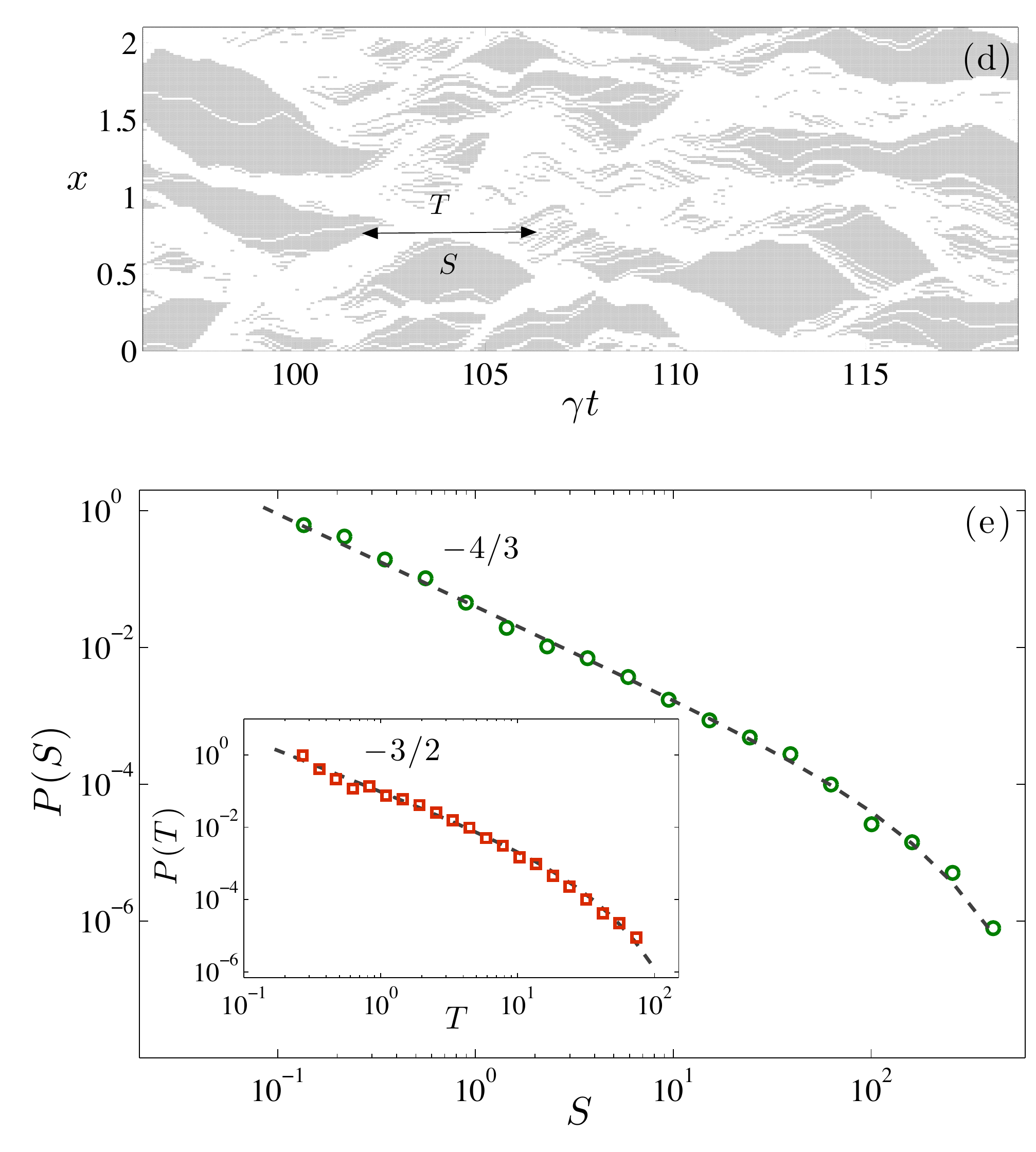}
\caption{(a) The distribution of the numbers of particles in
the trails is very broad, and is well approximated by a power-law in the small-mass limit.
(b, c) Plots of the probability $P({\cal N},\delta x)$ for finding ${\cal N}$ particles
in a cluster of size $\delta x$. There are power-law dependencies, with two
different exponents characterising the sparse (b) and dense (c)
regions. For this figure we used $\epsilon=1.75\epsilon_{\rm c}$.
(d, e) Probability distributions of the areas $A$ and lifetimes $T$ of voids which
are defined as the grey areas in panel (d).}
\label{fig: 3}
\end{figure*}

If $z(t)$ is negative, this implies that nearby trajectories are converging towards each other. 
Figure \ref{fig: 2}(a) compares the minimum value of the FTLE over a sample of $M$ 
trajectories, denoted as $z_\mathrm{min}(t)$, with the average value of $z(t)$, termed the 
Lyapunov exponent $\lambda$. 
The crucial condition for chaos is that $\lambda$ is positive. However, in our simulations
we find that the minimum value is negative up to a very long time,
indicating that some trajectories show very long periods of convergence. 
In section \ref{sec: 5} we argue that for a fixed number of
particles $M\gg 1$,  the minimal FTLE approaches $\lambda$ algebraically as $t\to \infty$ with
\begin{equation}
\label{eq: 3}
\lambda-z_{\rm min}(t)\sim \frac{f(M)}{\sqrt{t}},
\end{equation}
where $f(M)$ is a function which increases monotonically (but slowly - approximately
logarithmically) with $M$. Figure \ref{fig: 2}(a) shows the mean
value of the minimum FTLE for \eqref{eq: 1}, compared
with a fit proportional to $t^{-\alpha}$, where $\alpha$ is a power close
to $1/2$ (the parameters are the same as for Figure \ref{fig: 1}).

While the FTLE is negative, nearby trajectories are converging towards each other.
Equation \eqref{eq: 3}) implies that the FTLE can be, in principle,
made negative for arbitrarily
long times by increasing the number of trajectories. This indicates that the closest
approach of trajectories should decrease very rapidly as $M$ increases. This fact 
is illustrated in figure \ref{fig: 2}(b), where we show how the smallest separation $\delta x_{\rm min}$
between any trajectories of the flow illustrated in figure \ref{fig: 1} decreases as the number of
trajectories $M$ increases. Evaluating an ensemble average of $\delta x_{\rm min}$, we
find a power-law behaviour, $\langle \delta x_{\rm min}\rangle \sim M^{-\Gamma}$, 
for $10<M<20000$ with $\Gamma\approx 1.6$.

Figures \ref{fig: 2}(a) and \ref{fig: 2}(b) present evidence that the most strongly converging
trajectories lead to very high particle density. Figure \ref{fig: 2}(c) illustrates a
complementary aspect of this phenomenon, by showing that
converging regions occupy a large fraction of the phase space
of our model. The  cumulative PDF of $z$ is
very broad: even at time $\gamma t=51.2$
(time has been made dimensionless by using the damping rate in
\eqref{eq: 1}), the probability of $z$ being negative is as
high as $\approx 1/3$.

\section{Scale-free behaviour}
\label{sec: 4}

Figure \ref{fig: 1} shows evidence that the trajectories cluster into groups
which we term \lq trails'. In figures \ref{fig: 2}(a) and \ref{fig: 2}(b) we showed evidence
that there is an extremely broad distribution of density within these
trails, including regions of extremely strong convergence. We also see
evidence that the distribution of the numbers of trajectories in each trail is
very broad, and characterised by a power-law. Figure \ref{fig: 3}(a) shows the
probability distribution of the weights of trails for \eqref{eq: 1} for the
parameters used in figure \ref{fig: 1}: we plotted the distribution of the number 
of trajectories inside an interval of length $\Delta x=L/4000$. We find that
discrete models for particle trajectories, analogous to the Scheidegger river
model \cite{Sch67,Hub91}, also show a similar power-law distribution of trail weights,
indicating that this power-law is not a consequence of differential structure of the flow,
and is therefore independent of properties of the FTLE.

We have described power laws which characterise the dense regions of figure \ref{fig: 1}.
It is also of interest to understand the sparsely covered regions of this plot, and we
find evidence that lacunarity of this image is also characterised by power laws.
Let $P({\cal N},\delta x)$ be the probability that an interval of width $\delta x$ surrounding
a given trajectory contains ${\cal N}$ other trajectories. In figures \ref{fig: 3}(b,c) we
plot $P({\cal N},\delta x)$ versus ${\cal N}$, on doubly-logarithmic scales, for several
values of $\delta x$. The plots suggest that when $\delta x \ll \xi$, $P({\cal N},\delta x)$
has a power-law dependence upon ${\cal N}$
\begin{equation}
\label{eq: 4}
P({\cal N},\delta x)\sim {\cal N}^\beta
\end{equation}
with two different exponents, $\beta_1>0$ when ${\cal N}$ is below the position of the
peak at ${\cal N}_{\rm max}$, and a different exponent $\beta_2<0$ above the
peak. The exponents $\beta_1$ and $\beta_2$ depend upon $\epsilon$
(the coupling constant), but not upon $\delta x$ (interval width).
We find that the exponent $\beta_1$ approaches zero
as $\epsilon\to \epsilon_{\rm c}$: we used a larger value, $\epsilon=1.75\epsilon_{\rm c}$
in figures \ref{fig: 3}(b,c) so that $P({\cal N},\delta x)$ would show typical behaviour, with a clearly defined maximum.

As well as investigating the sparse regions of figure \ref{fig: 1}, we also
investigated the PDF of the sizes of the voids, where there are no trajectories.
Figures \ref{fig: 3}(d,e) show the definition of the area $A$ and lifetime $T$ of a void and
how they are statistically distributed. Both plots show clear evidence for
power laws at large values, with exponents $-4/3$ and $-3/2$ respectively
(again, we used the same parameters as for figure \ref{fig: 1}). These
exponents are readily explained by a model involving first passage processes.

It is well known that
dynamical systems may have attractors with a fractal measure (often
called \emph{strange attractors}), thus leading to fractal clustering in phase
space. This implies 
a power-law dependence of the mean number of trajectories
$\langle {\cal N}\rangle$ in a ball of radius $\delta x$ surrounding a given trajectory:
$\langle {\cal N}\rangle\sim \delta x^{D_2}$, where $D_2$ is a fractal dimension which is
known as the correlation dimension\cite{Gra+83}. The power-laws which we
have described, however, go beyond the fractal properties of strange
attractors: whereas
the fractal dimension describes the spatial structure of the most densely
occupied regions, \eqref{eq: 4} describes the probability
distribution of the \emph{amount} of material in a region, rather than its spatial
structure. In addition, the existence of more than one exponent
demonstrates that our approach uncovers new properties of the system.
Figures \ref{fig: 2}(b) and  \ref{fig: 3}(b,c) indicate
that the power-law distributions describe
the sparsely occupied regions, as well
as the dense regions. It is also interesting to note that the usual explanation for the fractal
structure of the strange attractor, involving stretching and folding in phase space, is not applicable
to this model\cite{Wil+12}.

\section{Theory for minimum FTLE} 
\label{sec: 5}

Here we present arguments which support equation (\ref{eq: 3}).  
The arguments are most transparently presented for 
one-dimensional maps. They are also applicable to the continuous models
in the main text, equations (\ref{eq: 1}) 
and (\ref{eq: 5}),  by considering the evolution over a finite time interval.
For a one-dimensional map 
$x_{n+1}=F_n(x_n)$, the finite-time Lyapunov exponent of a trajectory with 
initial position $x_0$ after $N$ iterations is 
\begin{equation}
\label{eq: S1}
z(x_0,N)=\frac{1}{N}\,\ln\,\left(\frac{\partial x_N}{\partial x_0}\right)
\ .
\end{equation}
If the trajectory reaches position $x_j(x_0)$ after $j$ steps, starting 
from initial position $x_0$, then (using the chain rule) 
$z(x_0,N)$ is a mean value of 
logarithms of gradients of the map along the trajectory:
\begin{equation}
\label{eq: S2}
z(x_0,N)=\frac{1}{N}\sum_{j=1}^N \ln\,\vert F'_j(x_j(x_0))\vert
\ .
\end{equation}
The Lyapunov exponent \cite{Ott02} is $\lambda=\lim_{N \to \infty} z(x_0,N)$.
We quantify the closest approaches of trajectories by considering 
the minimal value of the FTLE for a set of $M$ trajectories after $N$ iterations 
of the map. This will be denoted by $z_{\rm min}(N,M)$. For any fixed value of $M$, no 
matter how large, this quantity converges to $\lambda$ as $N\to \infty$. 

The determination of $z_{\rm min}(N,M)$ is a problem which combines the large-deviation 
principal with extreme-value statistics. Because the dynamics is assumed to be chaotic, 
the FTLE (as expressed in equation (\ref{eq: S2})) may be regarded as a
mean value of a sequence of random variables. The probability distribution 
of the FTLE can then be described by large deviation theory \cite{Fre+84,Tou09}, 
so that for large $N$ the probability density of $z$ has the asymptotic form
\begin{equation}
\label{eq: S3}
P(z)\sim \exp[-NJ(z)]
\end{equation}
where $J(z)$ is a function which is termed a \emph{rate function} 
or \emph{entropy function} \cite{Fre+84,Tou09}. If we take a fixed number of trajectories 
and consider the long-time limit, $N\to \infty$, the $M$ different trajectories may be 
assumed to be drawn independently from a probability density 
in the large deviation theory form, equation (\ref{eq: S3}). We are interested 
in the smallest value of $z$ for this sample of $M$ trajectories, $z_{\rm min}$.
This problem in extreme-value statistics can be addressed by the method 
introduced by Gumbel \cite{Gum35}. 
In order to make a rough estimate of $z_{\rm min}(N,M)$, it is sufficient 
to find the value of $z$ for which the exponential smallness of the probability 
density balances the large number of samples, $M$, that is 
\begin{equation}
\label{eq: S4}
M P(z_{\rm min})\sim 1
\ .
\end{equation}
In terms of the large deviation entropy function, this condition 
becomes: $M\,\exp[-NJ(z_{\rm min})]=1$. The logarithm of this 
equation gives the condition 
\begin{equation}
\label{eq: S5}
\ln M-NJ(z_{\rm min})=0
\ .
\end{equation}
Now consider how equation ({\ref{eq: 3}) 
follows from equation (\ref{eq: S5}). In the limit as 
$N\to \infty$, where $z_{\rm min}$ 
approaches $\lambda$, we are concerned with small 
values of $J(z)$, where the entropy can be approximated by a quadratic 
function:
\begin{equation}
\label{eq: S6}
J(z)=\frac {(z-\lambda)^2}{2\sigma^2}
\end{equation}
indicating that equation (\ref{eq: S5}) takes the form of equation (\ref{eq: 3}), 
with $f(M)=\sigma\sqrt{2\ln M}$. However, when we made a careful 
numerical investigation of the distribution of $z_{\rm min}(N,M)$, we found
that this expression does not give an accurate estimate for $f(M)$. In the following, 
we discuss our conclusions about the correct form for $f(M)$.

Firstly, we use the method introduced by Gumbel~\cite{Gum35} to determine the probability
density of the minimum value more precisely than \eqref{eq: S5}. Namely, 
we find that the PDF of $z_{\rm min}$ is approximated by
\begin{equation}
\label{eq: S7} 
\rho_{\rm min}(z)={\cal C}\,Z\,F(Y)
\end{equation}
where ${\cal C}$ is a normalisation constant, and
\begin{equation}
\label{eq: S8}
Z=\frac{(\lambda-z)\sqrt{N}}{\sigma}
\ ,\ \ \ 
Y=\frac{Z^2}{2}
+\ln\left(\sqrt{2\pi}|Z|\right)
-\ln\,M
\end{equation}
with
\begin{equation}
\label{eq: S9}
F(Y)=\exp[-(Y+\exp(-Y))]
\ .
\end{equation}
Equations (\ref{eq: S7}), (\ref{eq: S8}) and (\ref{eq: S9}) indicate that the typical size 
of $z_{\rm min}$ is of the form of equation (\ref{eq: 3}), 
where the function $f(M)$ 
is actually a generalised Lambert function rather than a logarithm. 
A numerical integration indicates
that the mean and variance of the minimum of the scaled variable $Z_{\rm min}$ are, 
respectively,
\begin{equation}
\label{eq: S10}
\langle Z_{\rm min}\rangle\approx \bar Z-\frac{0.41}{\sqrt{\ln\,M}}
\ ,\ \ \ 
{\rm Var}(Z_{\rm min})\approx \frac{0.85}{\ln\,M}
\end{equation} 
where $\bar Z$ satisfies $Y(\bar Z,M)=0$.

Now let us consider some numerical evidence 
on the applicability of the distribution defined by equations (\ref{eq: S7})-(\ref{eq: S9}).
In order to be able to make a thorough numerical study we examined 
a simplified version of the equation of motion (\ref{eq: 1}), 
in the form of a 
map termed the \emph{correlated random walk}~\cite{Wil+12}: 
\begin{equation}
\label{eq: S11}
x_{n+1}=x_n+f_n(x_n)
\end{equation}
where $f_n(x)$ are continuous and bounded random functions, 
drawn by independent sampling from an ensemble at each iteration.
This map is a generalisation of a random walk, and can be used
as a discrete model for advection of particles in a random flow\cite{Wil+12}. 

\begin{figure}[h!]
\includegraphics[width=0.4\textwidth]{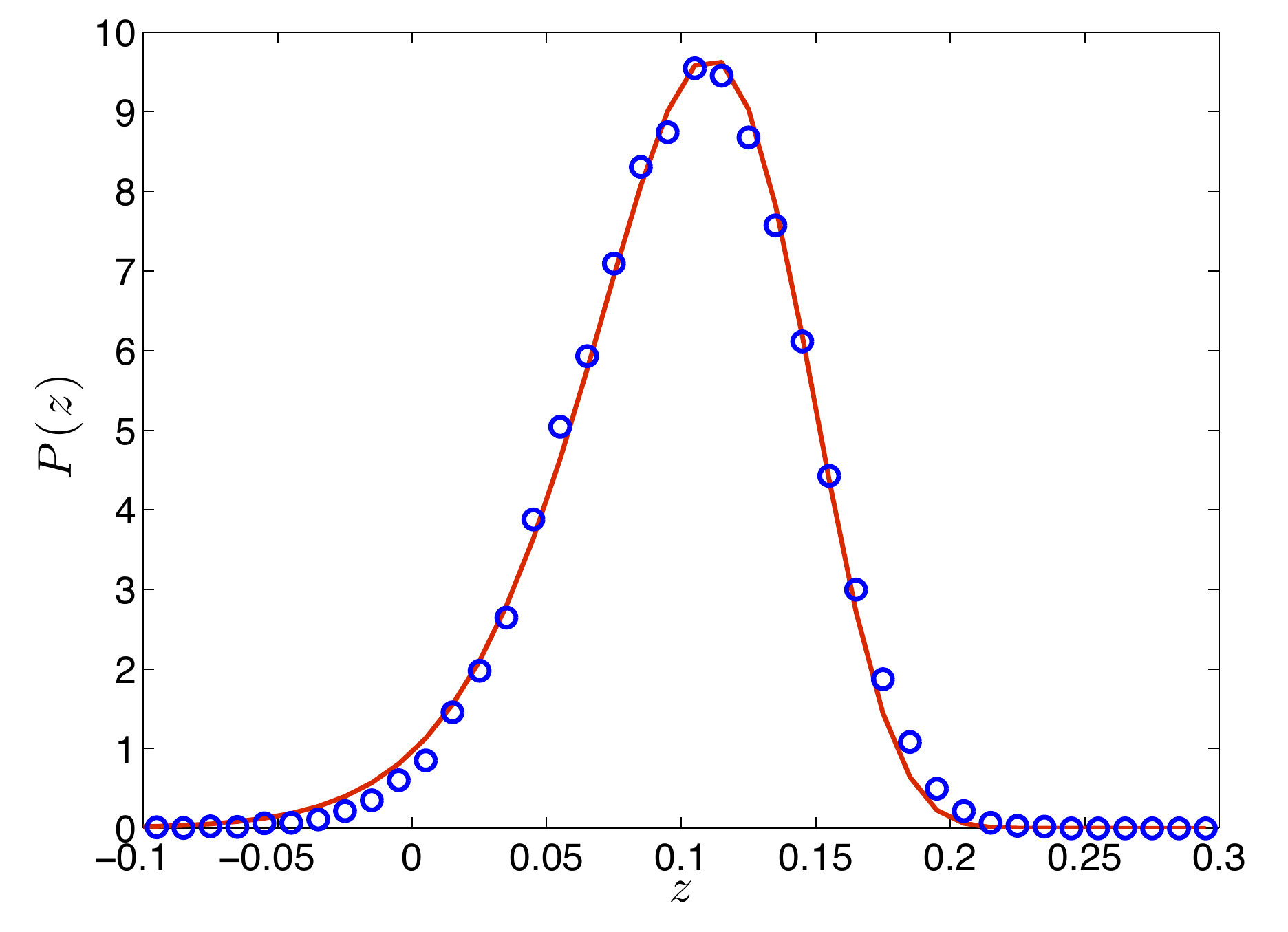}
\caption{
Probability distribution of the minimum FTLE for equation (\ref{eq: S11}), 
for a sample of $M=1000$ trajectories after $N=200$ iterations, accumulating results
using $2.5\times 10^4$ seeds of the random process in (\ref{eq: S12}).
This is compared 
with a fit to equations (\ref{eq: S7})-(\ref{eq: S9}): the effective
number of trajectories was $M_{\rm eff}=35.8$, and the variance $\sigma_{\rm eff}=1.2
\sigma$.}
\label{fig: S1}
\end{figure}

Our numerical studies considered the case where $f_n(x)$ has a Gaussian 
distribution, with the following statistics:
\begin{eqnarray}
\label{eq: S12}
\langle f_n(x)\rangle&=&0
\nonumber \\
\langle f_n(x)f_{n'}(x')\rangle&=&\varepsilon^2\xi^2\,\exp\left[-\frac{(x-x')^2}{2\xi^2}\right]\,\delta_{nn'}
\ .
\end{eqnarray}
We generated the $f_n(x)$ with approximately this correlation function 
by means of Fourier series, with period $L$ satisfying $\xi/L\ll 1$. The 
iterates $x_n$ are confined to the interval $[0,L]$ by adding an integer 
multiple of $L$ to $x_n$ every time a particle leaves the interval. This gives 
statistics which 
become stationary as $n\to \infty$. The quantities $\lambda$ and $\sigma$ are 
obtained from moments of the distribution of the gradient, $f'(x)$, which has a 
Gaussian distribution with variance $\varepsilon^2$.  
It is known that the Lyapunov exponent of this model, $\lambda=\langle |1+f'(x)|\rangle$, 
is positive for $\varepsilon>\varepsilon_{\rm c}$ with the critical point at 
$\varepsilon_{\rm c}=2.421\ldots$~\cite{Wil+12}. The numerical 
illustrations shown in figures \ref{fig: S1} and \ref{fig: S2}, 
were for the case $\varepsilon=1.5\,\varepsilon_{\rm c}$, where $\lambda\approx 0.302$ 
and $\sigma\approx 1.107$.

\begin{figure}[ht!]
\includegraphics[width=0.4\textwidth]{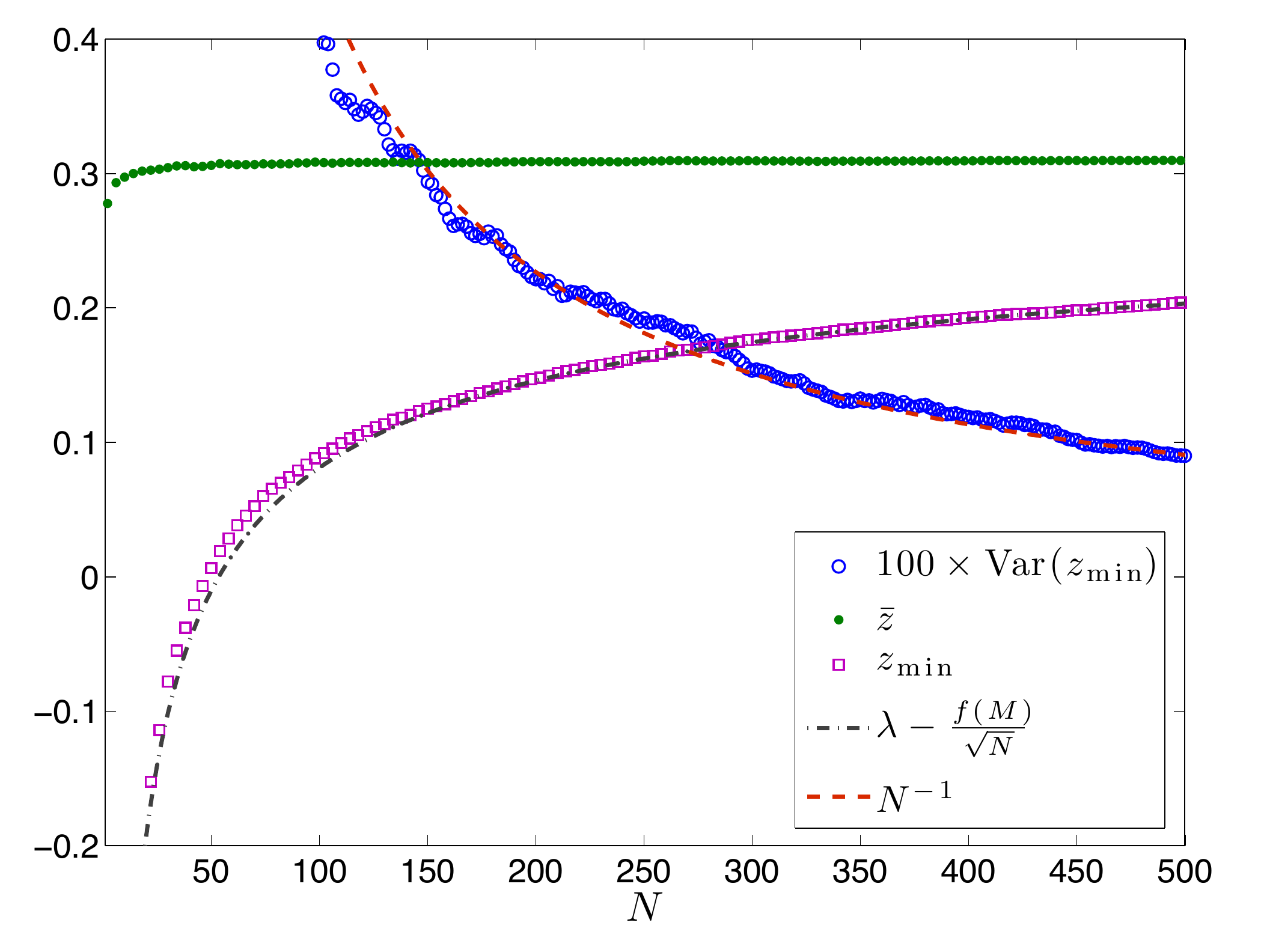}
\caption{
Average of the minimum value of the finite-time Lyapunov 
exponent over $10^3$ realisations of the random functions in 
the correlated random walk model, equations (\ref{eq: S11})-(\ref{eq: S12}).  
The data are plotted as a function of the number of iterations, $N$, and 
compared with equation (\ref{eq: 3}), 
where the used the fitting parameters 
$M_{\rm eff}=18.3$ and $\sigma_{\rm eff}=1.125\sigma$ in equations (\ref{eq: S10}). 
We also show the variance of $z_{\rm min}$, compared with the fitted value.
}
\label{fig: S2}
\end{figure}

We examined the probability distribution of $z_{\rm min}$, finding that it 
is of the form (\ref{eq: S7})-(\ref{eq: S9}), with 
$M$ and $\sigma$ replaced by effective values, $M_{\rm eff}$ and $\sigma_{\rm eff}$. 
We find $\sigma_{\rm eff}/\sigma \approx 1$, and the \small  
difference is likely to be a consequence 
of the fact that $J(Z)$ is only approximately quadratic. However we find that $M_{\rm eff}/M\ll 1$.
We interpret this as being a consequence of the clustering of trajectories illustrated in 
figure 1 of the main text. Because many of the trajectories are very closely clustered 
together, the number of independent samples of the phase space is much less than $M$.
Figure \ref{fig: S1} shows the probability distribution of $z_{\rm min}(N,M)$ for $M=10^3$ 
trajectories after $N=100$ iterations. 
There is an excellent fit to the distribution (\ref{eq: S7})-(\ref{eq: S9}), 
with $\sigma_{\rm eff}=1.15\sigma$ and $M_{\rm eff}=35.8$.

We computed an ensemble average over different realisations of the random 
functions in equation (\ref{eq: S12}).
The ensemble averaged results are shown in figure \ref{fig: S2}, which shows the mean
FTLE converging to $\lambda=0.302$, and the average of $z_{\rm min}(N,M)$ over 
$10^3$ realisations, for $M=100$ trajectories, compared to a fit of equation
(\ref{eq: 3}): 
there is excellent agreement with the prediction that 
$\lambda-\langle z_{\rm min}\rangle\sim N^{-1/2}$. 
We also computed the variance of $z_{\rm min}(N,M)$, which is  
asymptotic to a multiple of $N^{-1}$. Using the 
mean and the variance we were able to determine the 
two parameters $\sigma_{\rm eff}$ and $M_{\rm eff}$,
obtaining $M_{\rm eff}=18.3$ and $\sigma_{\rm eff}=1.125\sigma$. 
This allowed us to fit the data to equations (\ref{eq: S10}).
We repeated this for different
values of the number of trajectories, namely $M=100$, $10^3$ and $10^4$ 
trajectories, and we found fitted values of 
$\sigma_{\rm eff}/\sigma$ equal to $1.125$, $1.15$ and $1.175$ respectively. 
The fitted values of $M$ were $M_{\rm eff}=18.3$, $35.8$ and $75.5$ respectively. 
This is consistent with another power-law relation,
\begin{equation}
\label{eq: S13}
M_{\rm eff}=\mu M^\Gamma
\end{equation}
with $\Gamma\approx 0.30$ for $\varepsilon=1.5\varepsilon_{\rm c}$.

\section{Perpetually converging trajectories}
\label{sec: 6}

In section \ref{sec: 5} we emphasised the effects of the slow approach of $z_{\rm min}(N,M)$ 
towards $\lambda$ in the long-time limit, $N\to \infty$. For any given value
of the number of iterations $N$ (or altenatively, for any time $t$), equation 
(\ref{eq: 3}) indicates that $z_{\rm min}$ decreases as the number of trajectories 
$M$ increases. This raises the question as to what is the limit of $z_{\rm min}(N,M)$ as 
$M\to \infty$ for a fixed but large value of $N$. There will be a global minimum 
$\underbar z(N)$ after $N$ iterations, which can be located 
by taking a sufficiently large number of initial conditions. 
Because of the exponential sensitivity of chaotic systems to their initial conditions, we 
expect that the number of trajectories, ${\cal M}$, required to accurately locate the global minimum 
of $z$ is ${\cal  M}\sim K^N$, for some constant $K$. 
If we replace $M$ with ${\cal M}=K^N$ in 
equation (\ref{eq: S5}), we obtain an equation $\ln\,K=J(z_{\rm min})$, which is independent 
of $N$. This suggests that the limit of $z$ as $M\to \infty$ should approach a limit $\mu$ as 
$N\to \infty$. This is not, however, a compelling argument because the derivation of (\ref{eq: S5})
assumed that we take $M$ independent random samples of the distribution of $z$. If we 
increase $M$ so as to sample the entire phase-space, we cannot guarantee that the 
trajectories which yield extreme values are independent of each other. 

However, there are arguments based upon exactly solvable systems which support the 
hypothesis that the global minimum of $z$ after $N$ iterations approaches a limit $\mu$ which is 
independent of $N$ and distinct from $\lambda$. Consider first a deterministic 
one-dimensional dynamical system for which it is obvious that $\mu<\lambda$. 
This is the generalised tent map 
\begin{equation}
\label{eq: 23}
x_{n+1}=
\left\{\begin{array}{lll}
g_1 x_x &,& 0\le x_n<g_1^{-1}\cr
g_2(1-x_n) &,& g_1^{-1}<x_n\le 1
\end{array}\right.
\ .
\end{equation}
The gradients of the linear sections, $g_1$ and $-g_2$ satisfy a 
harmonic mean value constraint: $g_1^{-1}+g_2^{-1}=1$.
This is a piecewise linear map of the 
interval $[0,1]$ into itself. 
The Lyapunov exponent is 
\begin{equation}
\label{eq: 24}
\lambda =\frac{g_2\,\ln\,g_1+g_1\,\ln\,g_2}{g_1+g_2}
\ .
\end{equation}
The $N$-fold composition of the map has $2^N$ piecewise linear intervals.
If $g_1<g_2$, then the interval with the smallest FTLE is the first interval, for which the 
instability factor is $g_1^N$ and hence
\begin{equation}
\label{eq: 25}
\mu=\ln\,g_1
\end{equation}
so that $\mu<\lambda$ if $g_1<g_2$. This elementary 
example shows that the minimal FTLE may converge to 
a value which is different from the Lyapunov exponent. The value of $\mu$ must, however, 
be positive for this map.

In order to see an example where $\mu $ may be negative while $\lambda$ is positive, 
implying that there is always at least one trajectory which is convergent for all times, 
we consider an alternative dynamical system. This system has two variables,  
$x_N$ and $y_N$, specifying the state at every iteration. The 
variables $y_N$ are iterated according to a simple tent map, representing a Bernoulli shift: 
$y_{N+1}=2y_N\,{\rm mod}\,1$. The iteration of the $x_N$ variable depends upon two random 
independent identically distributed random variables, $a_{n,\pm}$: 
\begin{equation}
\label{eq: 26}
x_{n+1}=\left\{\begin{array}{lll}
a_{n,+}\times x_n&,& 0<y_n<\frac{1}{2} \cr
a_{n,-}\times x_n &,& \frac{1}{2}<y_n<1 
\end{array}\right.
\ .
\end{equation}
We draw the $a_{n,\pm}$ independently from the same probability distribution. 
The initial condition for the $y_N$ variables is $y_0=x_0$. 
Consider the dynamics generated by this process as a map
$x_0\rightarrow x_N$. The map is piecewise linear on a set 
of intervals, which are determined by the discontinuities of the 
process which generates the auxiliary variables $y_N$. After 
$N$ iterations there are $2^N$ such intervals. Within each interval, we have
\begin{equation}
\label{eq: 27}
x_N=\left[ \prod_{j=1}^N a_{j,\pm}\right]  x_0
\end{equation}
where the $a_{j,\pm}$ are the random variables selected at random at each 
iteration, either $a_{j,+}$ or $a_{j,-}$ depending upon the trajectory $x_j$. 
These variables are chosen independently for each of the $2^N$ 
intervals. The FTLE for a given trajectory is, therefore,
\begin{equation}
\label{eq: 28}
z=\frac{1}{N}\sum_{j=1}^N \ln\,|a_{j,\pm}|
\end{equation}
which is a mean value of a sum of random variables. 

\begin{figure}[h!]
\includegraphics[width=0.45\textwidth]{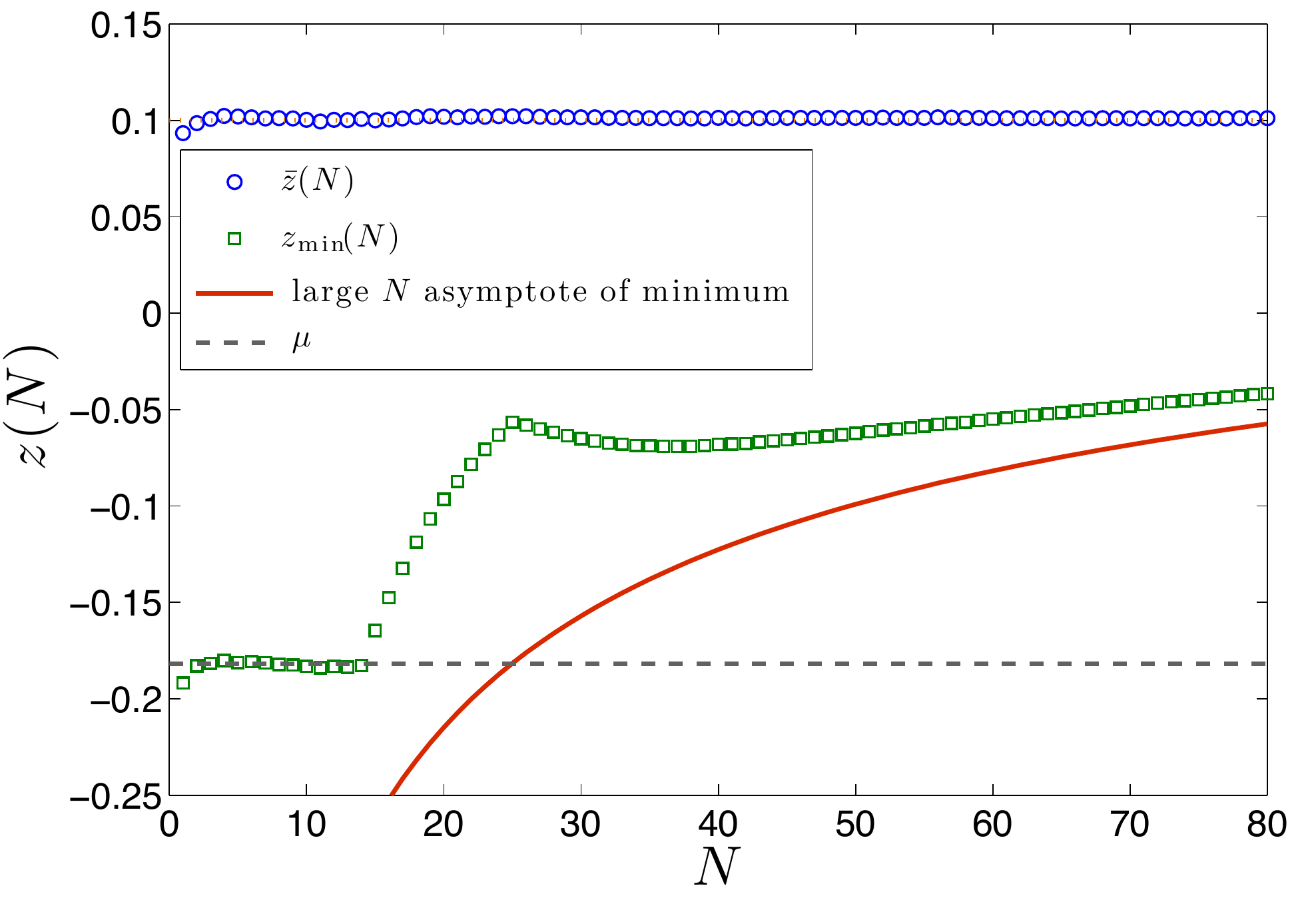}
\caption{
The ensemble average of the mean and minimum Lyapunov exponent for the 
map described by equations (\ref{eq: 26})-(\ref{eq: 28}). Here $\sigma_y=0.5$, $\lambda=0.1$ 
and $M=2^{15}$, so that the ensemble average of the minimum of $z$ is equal
to $\mu$ up until $N=14$. For large $N$ it approaches the asymptote given by equations 
(\ref{eq: 3}) and (\ref{eq: S6})-(\ref{eq: S10}), with $\sigma$ replaced by $\sigma_y/\sqrt{2}$. 
For this model, $M$ is not replaced by an effective value.}
\label{fig: 7}
\end{figure}

For the sake of definiteness, we make a simple and convenient choice 
for the statistics of the variables $a_{j,\pm}$. It is convenient to take the $a_{j,\pm}$ 
log-normally distributed, so that the probability distribution function of 
$y=\ln\,a$ is $P(y)=\exp[-(y-\lambda)^2/2\sigma_y^2]/\sqrt{2\pi}\sigma_y$, where $\lambda$ 
and $\sigma_y$ are two parameters. The variance of the sum is with respect to 
different choices of signs, but a fixed realisation of the $a_{j,\pm}$, is $\sigma_y^2/2$,
so that $\sigma=\sigma_{y}/\sqrt{2}$ in equations (\ref{eq: 3}), and (\ref{eq: S6})-(\ref{eq: S10}). 
The ensemble average of the minimum value of $z$ over all choices of signs is 
equal to the average of the smaller of $a_{j,+}$ and $a_{j,-}$, which is approximately 
$-0.564\sigma_y$. The ensemble average of the minimum of $z$ is expected to equal 
\begin{equation}
\label{eq: 29}
\mu \approx \lambda-0.564\sigma_y
\end{equation}
until $2^N>M$, at which point the trajectories do not explore phase space in sufficient 
detail to identify the global minimum of $z$. These prediction were verified by a numerical 
experiment (see figure \ref{fig: 7}).

\section{Applications}
\label{sec: 7}

\subsection{Particle concentration in surface flows}

We have used a one-dimensional model to illustrate our model, because it allows
us to represent the space-time structures of the trajectories in a two-dimensional
image such as Figure \ref{fig: 1}. There is, however, nothing in our discussion which is
specific to one dimension, and the three-dimensional version of our model \eqref{eq: 1}
is frequently used to describe the motion of particles in complex flows.
It is already known that turbulent flows can induce fractal particle clustering\cite{Som93},
although the effect is weaker than that illustrated in figure \ref{fig: 1}, because the underlying
fluid flow is incompressible~\cite{Bec07}
(whereas our one-dimensional model,
of necessity, involves a compressible flow).
Clustering effects are believed to play a role in the production of
raindrops in clouds~\cite{Sun+97,Pum+16} (note however that other effects may
be crucial to these processes~\cite{Kos+05,Wil16}).

The very strong convergence property, exhibited in Figure \ref{fig: 1} is
very reminiscent of the clustering of particles floating on the
surface of a turbulent water tank \cite{Lar+09}.
We remark that particles floating
on the surface of a turbulent fluid experience a compressible and apparently random flow field.
Experiments indicate that the
correlation function of the particle distribution is $C(\Delta r)\sim \Delta r^{-0.92\pm 0.02}$ \cite{Lar+09},
implying that the particles cluster with a correlation dimension $D_2\approx 0.08$ \cite{Gra+83}.
These observations show that surface flows are very close to a critical point at which path
coalescence occurs. We modelled a surface flow by the equations of motion
\begin{eqnarray}
\label{eq: 5}
\dot{\mbox{\boldmath$x$}}&=&\mbox{\boldmath$u$}(x,y,t),
\nonumber \\
\mbox{\boldmath$u$}&=&\mbox{\boldmath$\nabla$}\wedge\psi +\alpha \mbox{\boldmath$\nabla$}\phi,
\end{eqnarray}
where $\psi(x,y,t)$ and $\phi(x,y,t)$ are two independent, isotropic, homogeneous scalar fields
with a short correlation time, and $\alpha$ is an adjustable parameter.
In this case it is known that $D_2=2(1-\alpha^2)/(1+3\alpha^2)$ \cite{Bal+01},
so that we can model the flow by taking $\alpha = 0.926$. Figure \ref{fig: 8} shows a
simulation of this model for floating particles, where the particles become
concentrated along lines of convergence associated with sinking fluid.
Figure \ref{fig: 8} is very reminiscent of experimental images \cite{Lar+09},
validating the use of this model.

\begin{figure}[h t b]
\centering
\includegraphics[width=0.42\textwidth]{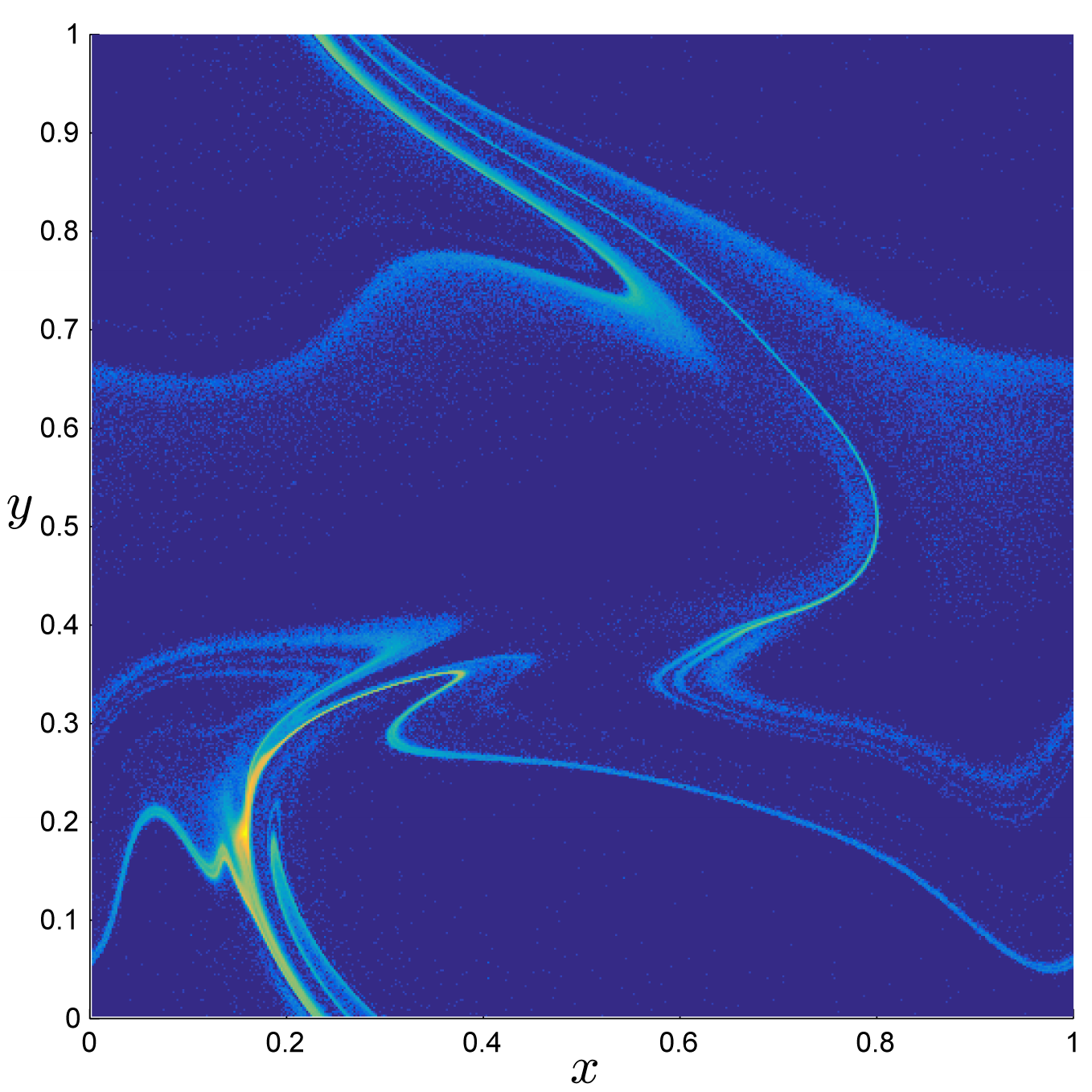}
\caption{Simulation of the distribution of particles floating over a complex two-dimensional
flow. (\eqref{eq: 5} with $\alpha=0.926$.)}
\label{fig: 8}
\end{figure}

The investigation of ~\eqref{eq: 5} in two spatial dimensions
reveals extreme convergence effects similar to those observed with
the one-dimensional model \eqref{eq: 1}.
This is illustrated by Figure \ref{fig: 8}, which shows the positions of $10^6$
particles, which were originally evenly distributed in
$10^6$ pixels. In Figure \ref{fig: 8}, one of the pixels has accumulated
nearly $3\times 10^4$ particles.
We have therefore provided evidence that, because surface flows are 
close to a critical
point, they exhibit a pronounced form of the convergence phenomena displayed in
Figure \ref{fig: 1}. 

We propose that these effects, which combine strong convergence
with mixing behaviour, may have played a role in the evolution of primitive living
organisms. Early organisms would have lacked the mobility required to follow
concentration gradients to find nutrients, to explore different environments, or
to encounter other individuals which might have advantageous mutations.
A process such as that illustrated in Figure \ref{fig: 8}, which combines
mixing and converging behaviours, seems to offer advantages to primitive
organisms. This supports the hypothesis that the first living organisms would have
evolved in the surface layers of water, and that motion of the water could act as a
catalyst for evolutionary development.

\subsection{Financial risks}

The arguments that we have presented are quite general, indicating
that the \emph{convergent chaos} phenomenon, involving transient 
convergence of chaotic trajectories may find applications in very different domains.
Insurance or futures transactions, where one
takes a fee in exchange for writing a contract which requires a payment 
to be made if there is a loss or an unfavourable change in the price,
may be an area ripe for the concept of convergent chaos. 
Substantial academic fields have developed around determining the 
value of these contracts. In insurance, actuarial methods are 
used \cite{Pro15}, and in finance, models based upon diffusive fluctuations of 
asset prices are the underlying tool \cite{Wil+95}. Any information about the 
nature of risk can be used to gain 
advantage. Our investigation shows that some chaotic systems, 
which would usually be assumed to be unpredictable, 
could be in fact highly predictable for certain initial conditions. 
Our results suggest that it may be 
possible to understand the conditions leading to a much smaller
uncertainty than expected, so that the risk in a futures contract would be 
reduced.

\section{Discussion}
\label{sec: 8}

Our results have shown that a simple chaotic dynamical system 
which describes the motion of particles in a turbulent flow can show an extremely 
high degree of convergence, despite the fact that the trajectories must eventually 
diverge with a positive rate of exponential growth. Using large-deviation and extreme-value 
concepts, we have shown that this transient convergence
may be very long-lived, intense and widespread (as illustrated by our studies of the finite-time 
Lyapunov exponent), and that it exhibits several scale-free geometrical properties, revealed 
by exhibiting power-law distributions. The \emph{convergent chaos} effect is expected to be 
observed in many systems, and we expect that it will be utilised for optimising the price 
of futures contracts. The model that we investigated in some depth, namely
motion of particles in a turbulent flow, shows particularly marked convergence in the case
of particles on the surface of a two-dimensional flow, and we argued that the combination 
of mixing and converging effects may have facilitated evolution of primitive organisms.

The phenomena described here have broad
implications for the interpretation of chaos, specifically of
the \lq butterfly effect'. Are perturbations destined to alter the course of large-scale
patterns in turbulent systems?  Or could regions of the phase space of a chaotic
dynamical system be screened off from small perturbations?
Our work clearly provides a positive answer to the latter question,
thus bringing new insight on the Lorenz' Brazilian butterfly problem. For these reasons
the converging divergence phenomenon is likely to lead to a deeper understanding of chaotic dynamics and of its
applications, and as such, deserves systematic investigation.

The authors are grateful to the
Kavli Institute for Theoretical Physics for support, where
this research was  supported in part by the National Science Foundation
under Grant No. PHY11-25915.

Author email addresses:

\indent \ \ \ marc.pradas@open.ac.uk
 
\indent \ \ \ alain.pumir@ens-lyon.fr

\indent \ \ \ huber@kitp.ucsb.edu

\indent \ \ \ m.wilkinson@open.ac.uk

\end{document}